\documentclass[superscriptaddress, aps,pre]{revtex4-2}
\usepackage{graphicx}
\usepackage{amsmath,amssymb}
\usepackage{graphics,color,array,calc,rotating,epsfig,psfrag}
\usepackage{hyperref}
\numberwithin{equation}{section}
\usepackage{bm}
\usepackage{dcolumn}
\usepackage{float}
\usepackage{subfigure}
\usepackage{amsfonts}
\usepackage[mathscr]{eucal}
\usepackage{helvet}
\def\be{\begin{equation}} \def\ee{\end{equation}}
\def\bea{\begin{eqnarray}} \def\eea{\end{eqnarray}}

\newcommand\prt{\partial}

\newcommand{\RN}[1]{%
  \textup{\uppercase\expandafter{\romannumeral#1}}%
}
\begin{document}
\baselineskip 18pt%
\begin{titlepage}
\vspace*{1mm}%
\hfill%
\vspace*{15mm}%
\hfill
\vbox{
    \halign{#\hfil         \cr
          } 
      }  
\vspace*{20mm}
\begin{center}
{\Large {\bf Thermodynamic Topology of Einstein-Maxwell-Dilaton Theories}}
 \end{center}
\vspace*{5mm}
\begin{center}
{H. Babaei-Aghbolagh,$^{\dagger}$ Habib Esmaili,$^{\pounds}$ Song  He,$^{\dagger,}$$^{\ddagger}$ and  Hosein Mohammadzadeh$^{\pounds}$ 
			}\\
			\vspace*{0.2cm}
			{\it
				$^{\dagger}$Institute of Fundamental Physics and Quantum Technology, \& School of Physical Science and Technology,  Ningbo University, Ningbo, Zhejiang 315211, China  \\
            $^{\pounds}$Department of Physics, University of Mohaghegh Ardabili,
				P.O. Box 179, Ardabil, Iran \\
                $^{\ddagger}$Max Planck Institute for Gravitational Physics (Albert Einstein Institute), Am M\"uhlenberg 1, 14476 Golm, Germany\\
			}

 \vspace*{0.5cm}
{E-mails: {\tt hosseinbabaei@nbu.edu.cn, h.esmaili@uma.ac.ir, hesong@nbu.edu.cn, mohammadzadeh@uma.ac.ir}}
\vspace{1cm}
\end{center}

\begin{abstract}
We present a systematic investigation of the thermodynamic topology for a broad class of asymptotically charged Anti-de Sitter (AdS) black holes in Einstein-Maxwell-Dilaton (EMD) theories, examining how scalar coupling parameters and spacetime dimensions influence black hole thermodynamics. Employing a topological approach that utilizes the torsion number of vector fields constructed from the generalized free energy, we characterize black hole states as topological defects within the thermodynamic parameter space.
Through analytical solutions spanning dimensions $d = 4$, $d=5$, and $d=6$, including the Gubser-Rocha model, we demonstrate that variations in the dilaton coupling constant $\delta$, particularly near its critical value $\delta_c$, induce transitions between distinct thermodynamic topological phases. Our analysis reveals that certain black hole solutions constitute a novel class designated as $W^{0-\leftrightarrow 1+}$, characterized by a torsion number $W = 1$ that corresponds to a unique stability structure. We establish that Gubser-Rocha models belong to this topological classification.
These results significantly expand the existing classification framework while reinforcing thermodynamic topology as a robust analytical tool for probing the universal properties of black holes in both gravitational and holographic contexts. The findings provide new insights into the relationship between microscopic couplings and macroscopic thermodynamic behavior in extended gravity theories.
\end{abstract}
\maketitle
\end{titlepage}

\section{Introduction}
Geometry and topology, as two fundamental branches of mathematics, play a crucial role in understanding physical phenomena. In theoretical physics, these branches serve not only as conceptual frameworks but also as powerful tools for practical applications. Notably, gravity itself is a geometric theory that describes the structure of spacetime. Beyond gravity, geometric methods have been instrumental in analyzing physical concepts such as the geometric phase in quantum mechanics and thermodynamic geometry, which has proven effective in studying the thermodynamics of various systems, including black holes \cite{misner1973gravitation,nakahara2018geometry,berry1984quantal,ruppeiner1995riemannian,provost1980riemannian,mirza2010thermodynamic,esmaili2024thermodynamic,keshavarzi2025thermodynamic}.

Topology, on the other hand, has also found extensive applications in physics. For example, topological insulators and topological phase transitions in condensed matter physics have garnered significant attention. These transitions describe quantum properties of materials characterized by the topology of their phase space. Furthermore, topology has been pivotal in understanding systems such as topological quantum field theories, stable light rings around black holes, and topological defect mappings in symmetric systems \cite{hasan2010colloquium,qi2011topological,witten1988topological,cunha2018shadows,cardoso2016gravitational,mermin1979topological}.

In recent years, the thermodynamics of black holes from a topological perspective has emerged as a compelling area of research, offering a novel framework for physicists. This approach leverages topological tools to analyze black holes as topological defects in the thermodynamic parameter space. Within this framework, black hole states can be classified by topological properties such as the winding number, which quantifies their topological characteristics. For instance, the stability or instability of black hole states can be directly correlated with the sign of their winding number \cite{wei2019repulsive,wei2022topology,kubizvnak2012p,chen2025thermodynamic,zhang2025topological}.
These recent advancements demonstrate that the application of geometric and topological tools not only deepens our understanding of fundamental concepts but also opens up new avenues for exploring physical systems and complex structures.

Thermodynamical topology provides a unified framework for classifying black hole thermodynamic behaviors, including phase transitions and critical points. The study introduces the concept of an "extended" thermodynamical topology that integrates existing topological approaches and establishes a connection between a black hole's critical exponents and its topological invariants\cite{wu2025extended}.

Black holes have long served as fundamental laboratories for exploring the interplay between gravity, thermodynamics, and quantum theory. Among the various approaches to studying these enigmatic objects, Einstein-Maxwell-Dilaton (EMD) theories have emerged as a powerful framework, offering a broad spectrum of analytic black hole solutions. These solutions, particularly those in asymptotically anti-de Sitter (AdS) spacetimes, are central to the AdS/CFT correspondence and provide a fertile ground for investigating the thermodynamic and topological properties of black holes \cite{gubser2001evolution,hirschmann2018black,chen2023descalarization,yerra2024topology,yerra2023topology}.

The thermodynamic topology of topological charged dilatonic black holes can be studied by treating them as topological defects, where the influence of dilaton gravity parameters and the topological constant on their thermodynamic properties allows classification into distinct topological classes according to their topological charges\cite{hazarika2024thermodynamic}.

This work focuses on a family of charged, asymptotically AdS$_{d+1}$ black holes supported by a neutral scalar field, parametrized by the spacetime dimension $d$ and a coupling constant $\delta$. A notable member of this family is the solution with $\delta = \delta_c \equiv \sqrt{\frac{2}{d(d-1)}}$, known as the Gubser-Rocha model. This solution has a top-down construction, arising from the near-horizon limit of D3-branes in ten-dimensional type IIB string theory for $d = 4$. These black holes exhibit rich thermodynamic behavior, making them ideal candidates for studying the connection between thermodynamics and topology \cite{gubser2009gravity,gubser2001evolution,ge2024thermo,lu2015thermodynamics, Babaei-Aghbolagh:2022xcy}.

The concept of thermodynamic topology has recently gained prominence as a novel method to classify black hole states. This framework views black holes as topological defects in a thermodynamic parameter space, where their properties can be characterized by winding numbers associated with the zero points of a vector field constructed from the gradient of the free energy. Each black hole state is thus endowed with a topological charge, which reflects its stability and role within the thermodynamic landscape. Locally stable black holes correspond to positive winding numbers, while locally unstable ones are associated with negative winding numbers. Summing the winding numbers over all zero points within a given parameter space yields a global topological number $W$, which serves as a classification parameter for the entire black hole system \cite{wei2022black,wei2022topology,fang2023revisiting,wu2023topological,chen2024thermodynamical,alipour2023topological}. Further, black hole thermodynamics in anti-de Sitter (AdS) spacetimes typically contains critical points in the phase diagram. One can classify these critical points topologically, including the one corresponding to the Hawking-Page transition as well~\cite{yerra2022topology,astefanesei2020dynamically, sadeghi2025phase,yerra2022topologyy,yerra2022topologyyy}.

Thermodynamic topology has revealed that black holes can be classified into distinct \textit{topological classes}, each associated with a unique global topological number. Previous studies have identified three such classes, corresponding to different configurations of black hole thermodynamic states. However, recent findings challenge this conventional classification, suggesting the existence of additional classes with unique topological features. These developments underscore the importance of topology in understanding the universal aspects of black hole thermodynamics and their underlying geometric structure \cite{wei2022black,wei2022topology,fang2023revisiting}.

In this work, we apply the framework of thermodynamic topology to analytically investigate the interior and thermodynamic properties of the family of charged AdS$_{d+1}$ hairy black holes arising from the EMD theory. By exploring the cases $\delta < \delta_c$, $\delta = \delta_c$, and $\delta > \delta_c$, we aim to uncover how the topological classification of black holes varies with the parameter $\delta$ and the spacetime dimension $d$, with a particular focus on the four-dimensional ($d = 4$) and five-dimensional ($d = 5$) cases \cite{gouteraux2014charge,arean2024kasner}. 

Our results reveal intricate connections between the scalar coupling, dimensionality, and the topological classification of black holes. We show that the thermodynamic topology not only provides a systematic classification of black hole solutions but also sheds light on their stability and phase transitions. In particular, the $\delta = \delta_c$ case, due to its origins in string theory, offers a compelling example of how top-down constructions can provide insights into the universal aspects of black hole topology and thermodynamics. Through this analysis, we extend the existing understanding of thermodynamic topology, enriching its role in the study of black hole interiors and their classification. This work contributes to the broader effort of connecting classical black hole physics with the geometric and topological insights essential to quantum gravity.

The outline of the paper is as follows: 
In section~\ref{2}, it introduces the concept of thermodynamic topology, where black holes are treated as topological defects in parameter space, classified by winding numbers that indicate stability and phase transitions. 
In section~\ref{3} presents the EMD models in detail, starting with a general family of asymptotically $\textbf{AdS}$ black holes and then focusing on specific cases: $\textbf{AdS}_4$ (Gubser-Rocha model), $\textbf{AdS}_5$, and $\textbf{AdS}_6$ black holes, each analyzed through their thermodynamic quantities, free energy, and topological classification using winding numbers.
Finally, the conclusions highlight the discovery that these black holes consistently fall into a new thermodynamic topological class $W^{0- \leftrightarrow 1+}$, revealing deeper insights into the universality of black hole thermodynamics and pointing toward future research directions.

\section{Thermodynamic Topology of Black Holes}\label{2}

Black holes serve as a unique testing ground for quantum gravity theories, exhibiting thermal properties with entropy proportional to the area of their event horizons \cite{bekenstein1973black,hawking1975particle,noori2025thermodynamic,gashti2025thermodynamicc}. Over the past decades, their thermodynamic framework has been extended to include concepts of pressure and volume \cite{creighton1995quasilocal,kastor2009enthalpy,cvetivc2011black}. This framework has provided deep insights into gravitational phase transitions \cite{hawking1983thermodynamics,wei2015insight} and holographic interpretations such as heat engines \cite{johnson2014holographic}, complexity \cite{al2021holographic}, and higher-dimensional theories \cite{frassino2023higher}.

Despite significant progress, the universal properties of black hole thermodynamics remain elusive. A promising approach to this challenge employs topology to classify black hole solutions as topological defects in the thermodynamic parameter space \cite{wei2022black,wei2024universal}. These classifications are based on the generalized off-shell free energy:
\begin{equation}\label{free-energy}
    \mathcal{F} =  \mathcal{E} - \frac{S}{\tau},
\end{equation}
where \(  \mathcal{E} \), \( S \), and \( \tau \) denote the energy, entropy, and cavity temperature of the black hole, respectively. By introducing an angular parameter \(\Theta \in (0, \pi)\), a vector field \(\phi\) is defined as:
\begin{equation}
    \phi = \left( \frac{\partial \tilde{\mathcal{F}}}{\partial r_h}, \frac{\partial \tilde{\mathcal{F}}}{\partial \Theta} \right),
\end{equation}
where \(\tilde{\mathcal{F}} = \mathcal{F} + \frac{1}{\sin\Theta}\). Black hole states correspond to the zero points of \(\phi\), characterized by a winding number \(w\), which is calculated using Duan's \(\phi\)-mapping topological current theory \cite{duan2000topological,duan2005non}. The total topological number, summing over all defects, is:
\begin{equation}
    W = \sum_{i=1}^N w_i,
\end{equation}
where \(N\) is the number of defects.

Recent studies have refined this framework into four thermodynamic topological classes based on the asymptotic behavior of the inverse temperature \(\beta = 1/T\) near the black hole's event horizon radius \(r_h\) \cite{wei2022black,wei2024universal}. The established classes, denoted \(W^{1-}\), \(W^{0+}\), \(W^{0-}\), and \(W^{1+}\), are defined as follows:
\begin{itemize}
    \item \textbf{Class \(W^{1-}\)}: \(\beta(r_m) = 0\) and \(\beta(\infty) = \infty\). Black holes in this class are unstable in both their small and large states.
    \item \textbf{Class \(W^{0+}\)}: \(\beta(r_m) = \infty\) and \(\beta(\infty) = \infty\). These black holes have stable small states and unstable large states.
    \item \textbf{Class \(W^{0-}\)}: \(\beta(r_m) = 0\) and \(\beta(\infty) = 0\). Black holes in this class have unstable small states and stable large states.
    \item \textbf{Class \(W^{1+}\)}: \(\beta(r_m) = \infty\) and \(\beta(\infty) = 0\). Black holes in this category are stable in both their small and large states.
\end{itemize}

The global topological number \(W\) serves as a classification criterion, derived from the sum of local topological numbers \(w\): \(w = +1\) for stable states and \(w = -1\) for unstable states. 

However, not all black holes fit within these four classes. For example, charged AdS black holes in gauged supergravity theories, such as four-dimensional static-charged AdS black holes in the Horowitz-Sen theory, or the five-dimensional Kaluza-Klein gauged supergravity theory, demonstrate thermodynamic transitions not captured by these classes \cite{wu2024topological}. Similarly, the dyonic AdS black hole exhibits distinct thermodynamic behaviors that preclude its inclusion in these categories \cite{chen2024thermodynamical}.

To address these gaps, new topological classes, \(W^{0-\leftrightarrow 1+}\), \(\overline{W}^{1+}\), and \(\hat{W}^{1+}\), have been proposed. These exhibit unique sequences of stability and instability transitions driven by temperature-dependent evolution. Black holes in these classes exhibit intricate thermodynamic behavior, particularly at low and high Hawking temperatures, which is distinct from previously established categories.
Also a newly identified subclass, $\widetilde{W}^{1+}$, reveals a unique stability profile with one stable small black hole at low temperatures and three coexisting states at high temperatures, necessitating an extension of the topological classification \cite{Ai:2025vno}.

In the following sections, we analyze these topological classifications in the context of the Einstein-Maxwell-Dilaton model. After reviewing the thermodynamic properties of the associated black holes, we explore their topology and investigate the implications of these new classes.

\section{Einstein-Maxwell-Dilaton models} \label{3}
\subsection{ Models: asymptotically AdS$_{d+1}$ family}
 We consider a simple extension of the Einstein-Maxwell theory, where the Maxwell field has an "exponential coupling" to an additional scalar field, known as the dilaton.
In this section, we consider the general bulk action of EMD theories in arbitrary dimensions, which is given by \cite{gouteraux2014charge}
\be\label{pact}
I = {1 \over 16\pi G_{d+1}} {\int_{}^{}} d^{d+1}x   \sqrt{-g}  \left[  R -  \frac{1}{2}(\partial_\mu\phi)^2 + V(\phi) -  \frac{1}{4} Z(\phi) F_{\mu\nu}^2    \right],
\ee
where $G_{d+1}$, $\phi(r)$, and $F_{\mu\nu}\!=\!\prt_{\mu}A_{\nu}-\prt_{\nu}A_{\mu}$ are the ${d+1}$-dimensional Newton's constant, dilaton field and field strength of $U(1)$ gauge field $A_{\mu}$, respectively. Also, $V(\phi)$ is a dilatonic scalar potential and $Z(\phi)$ is a coupling function between the Maxwell and dilaton fields.
 The terms denoted by $Z(\phi)$ and $V(\phi)$ are the coupling and potential function given as
 \begin{equation}\label{costant}
   Z(\phi)=e^{-(d-2)\delta\phi}, \qquad V(\phi)=V_{1}e^{\frac{(d-2)(d-1)\delta^2-2}{2(d-1)\delta}}+V_2e^{\frac{2\phi}{\delta-d\delta}}+V_3e^{(d-2)\delta\phi}.
 \end{equation}
 Here $V_1$, $V_2$ and $V_3$ are there constant parameters given by
 \begin{equation}\label{3-1}
     \begin{aligned}
         V_1&=\frac{8(d-2)(d-1)^3\delta^2}{(2+(d-2)(d-1)\delta^2)^2}, \qquad V_2=\frac{(d-2)^2(d-1)^2(d(d-1)\delta^2-2)\delta^2}{(2+(d-2)(d-1)\delta^2)^2},\\
         V_3&=-\frac{2(d-2)^2(d-1)^2\delta^2-4d(d-1)}{(2+(d-2)(d-1)\delta^2)^2},
     \end{aligned}
 \end{equation}
 where $\delta$ is a free parameter. It should be noticed that for $d=3$ by choosing $\delta=\sqrt{\frac{1}{3}}$ and redefinition $\phi\to -\sqrt{3}\phi$ the bulk action \eqref{pact} reduced to 4D EMD theory \cite{babaei2022complexity}, while for $d=4$ , $\delta=\sqrt{\frac{1}{6}}$ and redefinition $\phi\to -\sqrt{24} \phi$, it yields the 5D theory\cite{babaei2022complexity}.
 A metric for a higher-dimensional dilaton black hole including a cosmological constant has been developed, where the dilaton potential is shown to be a combination of three Liouville-type potentials and the cosmological constant is coupled to the dilaton in a non-trivial way\cite{gao2005higher}.
 The EMD theory, examined in relation to Eq.\eqref{3-1}, is consistent with the models reviewed in \cite{gao2005higher}.
 The equations of motion resulting from the action \eqref{pact} take the form
\begin{equation}\label{EOM}
    \begin{aligned}
        R_{\mu\nu}&=\frac{1}{2}\partial_{\mu}\phi\partial_{\nu}\phi+\frac{Z(\phi)}{2}F_{\mu}^{\rho}F_{\nu\rho}-\frac{Z(\phi)F^2}{4(d-1)}g_{\mu\nu}-\frac{V(\phi)}{d-1}g_{\mu\nu},\\
        0&=\nabla_{\mu}(Z(\phi)F^{\mu\nu}),\\
        0&=\square\phi + V'(\phi)-\frac{1}{4}Z'(\phi)F^2.
    \end{aligned}
\end{equation}
We are interested in solutions that realize homogeneous charged black hole geometries
(that is, solutions that depend only on the radial coordinate). These can be obtained via
the ansatz
\begin{equation}
    ds^2=-f(r) W(r)dt^2+\frac{W(r)}{f(r)}dr^2+U(r)^2 W(r)d\vec{x}^2_i,\qquad A=A_t(r)dt,\qquad \phi=\phi(r).
\end{equation}
Indeed, for the potentials \eqref{costant} one finds the following analytic solution of the equations \eqref{EOM}. The metric functions read
\begin{eqnarray}
 f(r)&=&\frac{r^2}{L^2} \Big( h(r)^{\frac{4 (d-1)}{(d-2) \bigl(2 + (d-2) (d-1) \delta^2\bigr)}}- \frac{r_h^d}{r^d}h(r_h)^{\frac{4 (d-1)}{(d-2) \bigl(2 + (d-2) (d-1) \delta^2\bigr)}} \Big) +(1-\frac{r_h^{d-2}}{r^{d-2} } )\kappa\\
 W(r)&=&h(r)^{\frac{6 - 2 d}{(d-2) \bigl(2 + (d-2) (d-1) \delta^2\bigr)}},\qquad U(r)=r\, h(r)^{ \frac{  d-1}{(d-2) \bigl(2 + (d-2) (d-1) \delta^2\bigr)}},\nonumber
\end{eqnarray}
where $r_h$ is the event horizon satisfying $f(r_h) = 0$. While the solutions for the matter fields are given by
\begin{equation}
    \begin{aligned}
    A_t(r)=&\frac{2 \sqrt{(d -1) (d-2) Q}  \left(1- \frac{r_h^{d -2}}{r^{d-2 }}\right) \sqrt{\frac{ \kappa  r_h^{d }}{ h(r_h)}+ r_h^{d +2} h(r_h)^{\frac{2 \left(2-\delta ^2 (d -2)^2 (d
   -1)\right)}{(d -2)  \left(\delta ^2 (d -2) (d -1)+2\right)}}}}{L\,(d -2) r_h^{d-1 } \sqrt{\delta ^2 (d -2) (d -1)+2} h(r)},\\
   e^\phi=&h(r)^{\tfrac{-2(d-1)\delta}{2+(d-2)(d-1)\delta^2}},\qquad h(r)=1+\frac{Q}{r^{d-2}}.
    \end{aligned}
\end{equation}
It is important to note that model \eqref{pact}-\eqref{costant} exhibits a dependence on two distinct parameters: the dimension $d$, and a real number $\delta$ which influences the functional forms of the potentials as defined in Eq.~\eqref{costant}.
 Finally, we shall point out that for the particular value of $\delta$
 \begin{equation}\label{deltacritical}
     \delta=\delta_c\equiv\sqrt{\frac{2}{d(d-1)}}.
 \end{equation}
The  temperature $(T)$,  entropy density $(s)$,  chemical potential $(\mu)$, and  charge density $(\rho)$ of the black hole are obtained as follows \cite{Arean:2024pzo}:
  \begin{equation}\label{TT1}
     T=h(r_h)^{\frac{1}{\frac{  d-2}{d-1} + \frac{1}{2} (d-2)^2 \delta^2}} \frac{r_h }{4 L^2 \pi}\Bigl(d -  \frac{4 ( d-1) Q}{h(r_h)\,  r_h^{d-2} \bigl(2 + ( d-2) (d-1) \delta^2\bigr)}\Bigr) + \frac{(d-2) h(r_h)^{\frac{2 - 2 d}{(d-2) \bigl(2 + (d-2) (d-1) \delta^2\bigr)}} \kappa}{4 \pi r_h},
 \end{equation}
 \begin{equation}\label{ss1}
     S=4 \pi  r_h^{d-1} h(r_h)^{\frac{1}{ \frac{d-2}{d-1} + \tfrac{1}{2} (d-2)^2 \delta^2}},
 \end{equation}
 \begin{equation}\label{mmuu}
     \mu=\frac{2 \sqrt{(d-2) (d-1) \, Q}  r_h^{1 - d}  \sqrt{h(r_h)^{-2 + \frac{4 (d-1)}{(d-2) \bigl(2 + (d-2) (d-1) \delta^2\bigr)}} r_h^{2 + d} + \frac{r_h^d \kappa}{h(r_h)}}}{L\,(d-2)  \sqrt{2 + (d-2) (d-1) \delta^2}},
 \end{equation}
 \begin{equation}\label{rrhhoo}
    \rho=\frac{2 \sqrt{(d-2) (d-1) \, Q} h(r_h)  \sqrt{h(r_h)^{-2 + \frac{4 (d-1)}{(d-2) \bigl(2 + (d-2) (d-1) \delta^2\bigr)}} r_h^{2 + d} + \frac{r_h^d \kappa}{h(r_h)}}}{L \,r_h \sqrt{2 + (d-2) (d-1) \delta^2}}.
 \end{equation}
In~\cite{yekta2021holographic,babaei2022complexity, Arean:2024pzo}, holographic complexity and its time evolution were analyzed for three families of holographic Einstein-Maxwell-Dilaton (EMD) theories with momentum relaxation driven by axionic scalar fields.
The free energy density $ \mathcal{E}$ in our model can be derived from the thermodynamic energy density $\mathcal{E}$ through the relation\cite{Arean:2024pzo,caldarelli2017phases}:
\begin{equation}
    \mathcal{E} = \frac{(d-1)}{d}\, (ST + \mu\rho).
    \label{eq:energy}
\end{equation}
Substituting the temperature, entropy density,  chemical potential, and charge density from Eqs.~\eqref{TT1}, ~\eqref{ss1},~\eqref{mmuu}, and ~\eqref{rrhhoo}, respectively, we obtain the energy density:
 \begin{equation}
      \mathcal{E}=(d-1) \left(\frac{r_h^d }{L^2} \, h(r_h)^{\tfrac{4 (d-1)}{(d-2) \left((d-2) (d-1) \delta ^2+2\right)}}+ 
   \left(\frac{4 (d-1) Q}{d \, L^2((d-2) (d-1) \delta ^2+2)}+\frac{(d-2)}{d} r_h^{d-2}\right) \kappa\right).
 \end{equation}

We begin our analysis by defining a generalized free energy function, introduced in Eq.\eqref{free-energy}.
For a black hole thermodynamic system with energy $\mathcal{E}$ and entropy $S$, where $\tau$ represents an additional variable analogous to the inverse temperature of a cavity surrounding the black hole.
Only when $\tau=1/T$ does the generalized Helmholtz free energy become on-shell.
A vector field $\phi$ is defined from $\mathcal{F}$ in the following way:
\begin{equation}
\phi=\left(\phi^{r_h},\phi^\Theta\right)=\left(\frac{\partial\mathcal{F}}{\partial r_h},-\cot{\Theta}\csc{\Theta}\right),
\end{equation}
 in which the two parameters $r_h$ and $\Theta$ oboy $0<r_h<+\infty$ and $0\leq\Theta\leq\pi$, respectively.
 For the Gubser-Rocha models-$\text{AdS}_{d+1}$ black hole, one can define the generalized Helmholtz free energy as
 \begin{equation}
     \begin{aligned}
         \mathcal{F}&=\frac{(d-1) r_h^d \left(Q r_h^{2-d}+1\right){}^{\tfrac{4 (d-1)}{(d-2) \left((d-2) (d-1) \delta ^2+2\right)}}}{L^2}+\frac{(d-1) \kappa  \left((d-2)
         r_h^{d-2}+\frac{4 (d-1) Q}{(L^2 (d-2) (d-1) \delta ^2+2)}\right)}{d}
         \\&-\frac{4 \pi  r_h^{d-1} \left(Q r_h^{2-d}+1\right){}^{\frac{1}{ \frac{d-2}{d-1} + \tfrac{1}{2} (d-2)^2 \delta^2}}}{\tau }\,.
     \end{aligned}
 \end{equation}
 The components of the vector $\phi$ can easily be calculated as:
 \begin{equation}
     \begin{aligned}
     \phi^{r_h}&=\frac{d (d-1) h(r_h)^{\frac{4 (d-1)}{(d-2) \bigl(2 + (d-2) (d-1) \delta^2\bigr)}} r_h^{ d-1} \Bigl(1 -  \frac{4 (d-1) Q r_h^2}{d h(r_h) r_h^d \bigl(2 + (d-2) (d-1) \delta^2\bigr)}\Bigr)}{L^2} \\
     &+ \frac{(d-2)^2 (d-1) r_h^{ d-3} \kappa}{d} -  \frac{4 (d-1) h(r_h)^{\frac{1}{- \frac{2 -  d}{d-1} + 1/2 (d-2)^2 \delta^2}} \pi r_h^{d-2} \Bigl(1 -  \frac{2 Q r_h^2}{h(r_h) r_h^d \bigl(2 + (d-2) (d-1) \delta^2\bigr)}\Bigr)}{\tau},
   \\ \phi^{\Theta}&=-\cot{\Theta}\csc{\Theta},
   \end{aligned}
 \end{equation}
 By solving the equation $\phi^{r_h}=0$, one can obtain a curve on the $r_h-\tau$ plane. For the d-dimensional Gubser-Rocha models, one can obtain
 \begin{equation}
    \begin{aligned}
        \tau=\frac{4 d h(r_h)^{\frac{2 (1 + d)}{(d-2) \bigl(2 + (d-2) (d-1) \delta^2\bigr)}} L^2 \pi r_h \Bigl(1 + \frac{(d-2) (d-1) Q r_h^2 \delta^2}{r_h^d \bigl(2 + (d-2) (d-1) \delta^2\bigr)}\Bigr)}{d^2 h(r_h)^{\frac{4 d}{(d-2) \bigl(2 + (d-2) (d-1) \delta^2\bigr)}} r_h^2 \Bigl(1 + \frac{(d-2) Q r_h^2 \bigl(-2 + (d-1) d \delta^2\bigr)}{d r_h^d \bigl(2 + (d-2) (d-1) \delta^2\bigr)}\Bigr) + (d-2)^2 h(r_h)^{1 + \frac{4}{(d-2) \bigl(2 + (d-2) (d-1) \delta^2\bigr)}} L^2 \kappa}.
    \end{aligned}
 \end{equation}

\subsection{$\textbf{AdS}_4$ black hole}
\begin{figure}[H]
        \centering
        \includegraphics[width=0.5\linewidth]{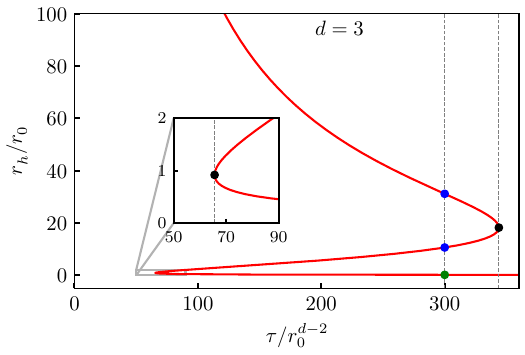}
        \caption{Zero points of the vector $\phi^{r_h}$ shown in the $r_h-\tau$ plane.}
        \label{fig1}
    \end{figure}

\begin{figure}[H]
    \centering
    \includegraphics[width=0.47\linewidth]{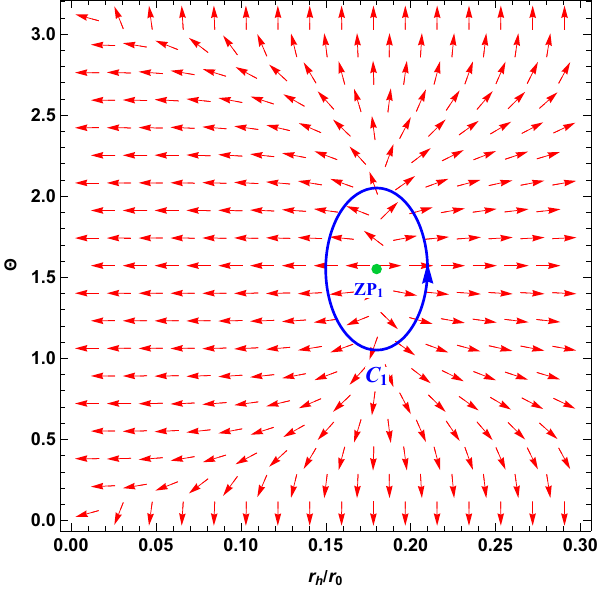}~~~
    \includegraphics[width=0.46\linewidth]{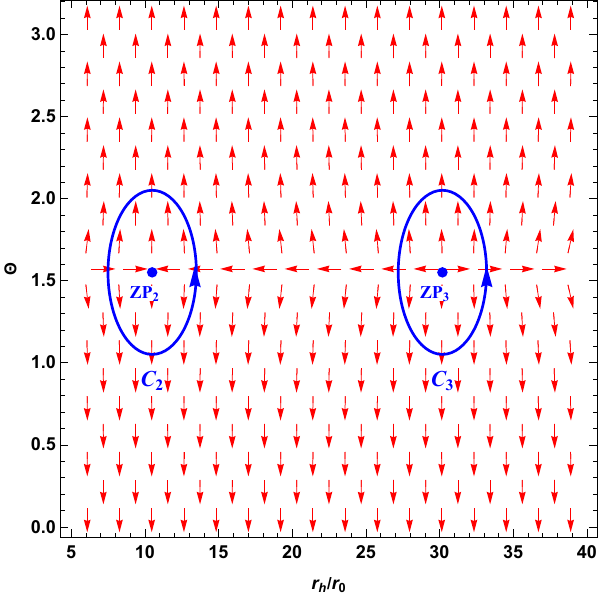}
    \caption{The unite vector field $n$ with $\tau=300r_0$ and $\kappa=1$.
    The zero points (ZPs) marked with green dot and blue dots at $(r_h,\Theta)=(0.18,\dfrac{\pi}{2})$ , $(r_h,\Theta)=(10.63,\dfrac{\pi}{2})$ and $(r_h,\Theta)=(31.15,\dfrac{\pi}{2})$, for $\text{ZP}_1$, $\text{ZP}_2$ and $\text{ZP}_3$, respectively.
    The blue contours $C_i$ are closed loops enclosing the zero points.}
    \label{fig2}
\end{figure}

In four dimensions, the EMAD theory consists of two main components: an EMD theory derived from the dimensional reduction of $\textbf{AdS}_4\times S^7$  in M-theory to the triple intersection of $M5$-branes, which yields the three-equal-charge black hole \cite{Kim:2017dgz}, and a kinetic term associated with two massless axionic scalar fields $\psi_I$ with $I=1,2$ \cite{gouteraux2014charge}.
In the $\textbf{AdS}_4$ black hole, the important quantities, including energy and entropy, used in calculating free energy $\mathcal{F}$ are as follows:
\begin{equation}
    \begin{aligned}
   \mathcal{E}&=\frac{2 r_h^3 \left(\frac{Q+r_h}{r_h}\right)^{\frac{4}{\delta ^2+1}}}{L^2}+\frac{2}{3} \kappa  \left(\frac{4 Q}{\left(\delta ^2+1\right)
   L^2}+r_h\right),\\
   S&=4 \pi  r_h^2 \left(\frac{Q+r^{}_h}{r^{}_h}\right)^{\frac{2}{\delta ^2+1}}.
    \end{aligned}
\end{equation}
For the Gubser-Rocha-AdS4 black hole, one can define the generalized Helmholtz free energy as
\begin{equation}
    \mathcal{F}=\frac{2 r_h^3 \left(\frac{Q+r_h}{r_h}\right)^{\frac{4}{\delta ^2+1}}}{L^2}+\frac{2}{3} \kappa  \left(\frac{4 Q}{\left(\delta ^2+1\right)
   L^2}+r_h\right)-\frac{4 \pi  r_h^2 \left(\frac{Q+r^{}_h}{r^{}_h}\right)^{\frac{2}{\delta ^2+1}}}{\tau }.
\end{equation}
The components of the vector $\phi$ can be easily calculated as:
\begin{equation}
     \begin{aligned}
     \phi^{r_h}&=\frac{2 \kappa }{3}+\frac{2 r_h \left(3 \delta ^2 (Q+r_h)-Q+3 r_h\right)
   \left(\frac{Q+r_h}{r_h}\right)^{\frac{4}{\delta ^2+1}-1}}{\left(\delta ^2+1\right)
   L^2}-\frac{8 \pi  \left(\delta ^2 (Q+r_h)+r_h\right)
   \left(\frac{Q+r_h}{r_h}\right)^{\frac{2}{\delta ^2+1}-1}}{\left(\delta ^2+1\right)
   \tau },
   \\ \phi^{\Theta}&=-\cot{\Theta}\csc{\Theta},
   \end{aligned}
 \end{equation}
 Solving the equation $\phi^{r_h}=0$ yields a curve in the $r_h-\tau$ plane. For four-dimensional Gubser-Rocha models, this results in:
  \begin{equation}
    \begin{aligned}
    \tau=\frac{12 \pi  L^2 r_h \left(\frac{Q+r_h}{r_h}\right)^{\frac{2}{\delta ^2+1}}
   \left(\delta ^2 (Q+r_h)+r_h\right)}{\left(\delta ^2+1\right) \kappa  L^2
   (Q+r_h)+3 r_h^2 \left(3 \delta ^2 (Q+r_h)-Q+3 r_h\right)
   \left(\frac{Q+r_h}{r_h}\right)^{\frac{4}{\delta ^2+1}}}.
    \end{aligned}
 \end{equation}
By analyzing the behavior of $r_h/r_0$ with respect to $\tau/r_0$ for different values of $\kappa$ ($\kappa = 0, 1, -1$), we can gain insights into their relations.
For a fixed dimensionless Radius length \textbf{AdS} of $L^2=500d(d-1)r_0^{d-1}$, where $r_0$ defines a characteristic length scale associated with the cavity enclosing the black hole, we examine the roots (zeroes) of $\phi^{r_h}$ Fig.~\ref{fig1}, and the unite vector field $n$ in Fig.~\ref{fig2} with $\tau=300r_0$ and $\kappa=1$.
Note that for these values of $L\sqrt{r_0^{1-d}}$ and $Q(\dfrac{r_h}{r_0})^{d-2}$, one generation point and one annihilation point can be found in Fig.\ref{fig1} at $\tau/r_0^{d-2}=65.56$ and $343.88$, respectively.
Calculating the winding number $W$ for these three branches of the black hole, according to Fig.\ref{fig1}, it can be seen that this behavior should be included in the new classifications presented in \cite{wu2025novel}.
The $\textbf{AdS}_4$ black hole is in the new thermodynamic topological classes \(W^{0-\leftrightarrow 1+}\), with the winding number of $1$.
 
\begin{figure}[H]
        \centering
        \includegraphics[width=0.5\linewidth]{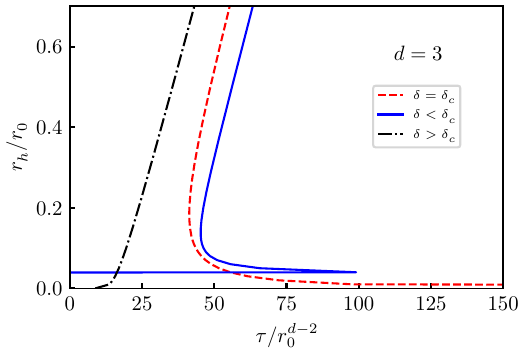}
        \caption{Zero points of the vector $\phi^{r_h}$ shown in the $r_h-\tau$ plane for $\delta=\delta_c~,~ \delta<\delta_c ~~\text{and}~~\delta>\delta_c$.}
        \label{fig3}
    \end{figure}

I consider a particular value for $\delta$ according to equation \eqref{deltacritical}, which is also called the $\delta_c$. Now, if we want to plot the graph of $r_h/r_0$ versus $\tau/r_0^{d-2}$ for values greater and less than the $\delta_c$, we can see the change in behavior for $\delta=\delta_c~,~ \delta<\delta_c ~~\text{and}~~\delta>\delta_c$ in Fig.~\ref{fig3}.
\subsection{ $\textbf{AdS}_5$ black holes}\label{sec32}
\begin{figure}[H]
        \centering
        \includegraphics[width=0.5\linewidth]{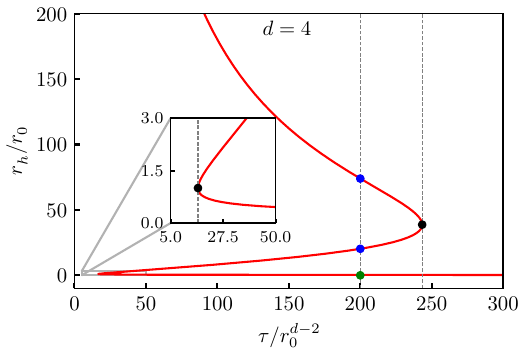}
        \caption{Zero points of the vector $\phi^{r_h}$ shown in the $r_h-\tau$ plane.}
        \label{fig4}
    \end{figure}

\begin{figure}[H]
    \centering
    \includegraphics[width=0.47\linewidth]{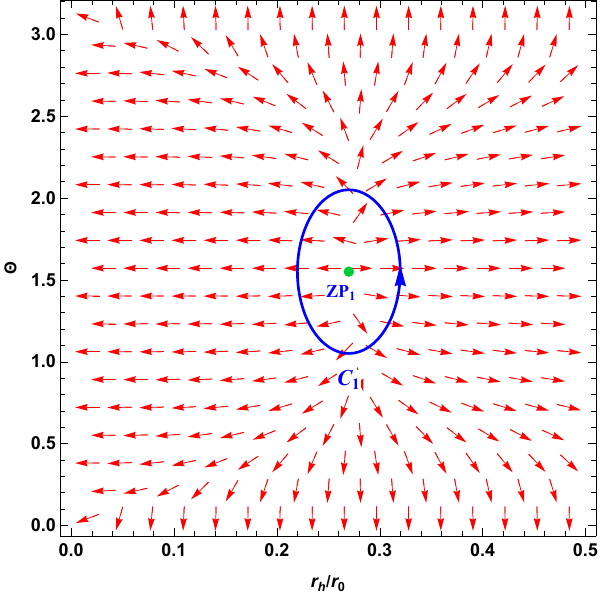}~~~
    \includegraphics[width=0.47\linewidth]{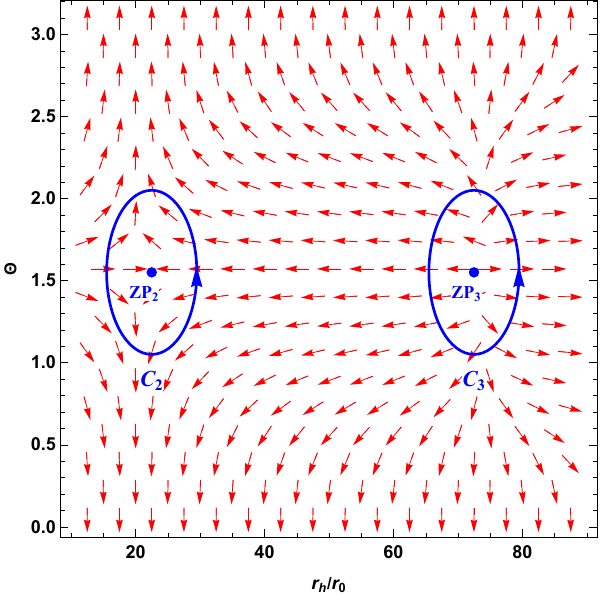}
    \caption{The unite vector field $n$ with $\tau=200r_0$ and $\kappa=1$.
    The zero points (ZPs) marked with green dot and blue dots at $(r_h,\Theta)=(0.27,\dfrac{\pi}{2})$ , $(r_h,\Theta)=(22.50,\dfrac{\pi}{2})$ and $(r_h,\Theta)=(72.50,\dfrac{\pi}{2})$, for $\text{ZP}_1$, $\text{ZP}_2$ and $\text{ZP}_3$, respectively.}
    \label{fig5}
\end{figure}

The generalization of the GR model \cite{Gubser:2009qt} to five-dimensional EMD theories was introduced in Ref.~\cite{gouteraux2014charge}. The EMD sector arises from 10-dimensional type IIB string theory, where the near-horizon geometry of D3-branes is given by $AdS_5\times S^5$ \cite{Cvetic:1999xp}.
The holographic charge transport and linear resistivity for charged $AdS_5$ black holes in this model have been reviewed in Refs.~\cite{gouteraux2014charge,jeong2018linear}. 
The energy of the black hole is given by
\begin{equation}
    \mathcal{E}=\frac{3}{2} \kappa  \left(\frac{3 Q}{\left(3 \delta ^2+1\right) L^2}+r_h^2\right)+\frac{3
   r_h^4 \left(\frac{Q}{r_h^2}+1\right)^{\frac{3}{3 \delta ^2+1}}}{L^2},
\end{equation}
On the other hand, the entropy of the black hole is
\be \label{ST5}
\begin{aligned}
    S=4 \pi  r_h^3 \left(\frac{Q}{r_h^2}+1\right)^{\frac{3}{6 \delta ^2+2}}.
\end{aligned}
\ee
Let us now consider the generalized free energy
\begin{equation}
    \mathcal{F}=\frac{3}{2} \kappa  \left(\frac{3 Q}{\left(3 \delta ^2+1\right) L^2}+r_h^2\right)+\frac{3
   r_h^4 \left(\frac{Q}{r_h^2}+1\right)^{\frac{3}{3 \delta ^2+1}}}{L^2}-\frac{4 \pi  r_h^3 \left(\frac{Q}{r_h^2}+1\right){}^{\frac{3}{6 \delta ^2+2}}}{\tau },
\end{equation}
The components of vector $\phi$ are given by:
\begin{equation}
     \begin{aligned}
     \phi^{r_h}&=\frac{6 r_h^3 \left(6 \delta ^2 \left(Q+r_h^2\right)-Q+2 r_h^2\right)
   \left(\frac{Q}{r_h^2}+1\right)^{\frac{3}{3 \delta ^2+1}}}{\left(3 \delta ^2+1\right) L^2
   \left(Q+r_h^2\right)}-\frac{12 \pi  r_h^2 \left(3 \delta ^2
   \left(Q+r_h^2\right)+r_h^2\right) \left(\frac{Q}{r_h^2}+1\right)^{\frac{3}{6
   \delta ^2+2}}}{\left(3 \delta ^2+1\right) \tau  \left(Q+r_h^2\right)}+3 \kappa 
   r_h,
   \\ \phi^{\Theta}&=-\cot{\Theta}\csc{\Theta},
   \end{aligned}
 \end{equation}
 Solving the equation $\phi^{r_h}=0$ yields a curve in the $r_h-\tau$ plane. For five-dimensional Gubser-Rocha models, this results in:
 \begin{equation}
    \begin{aligned}
        \tau=\frac{4 \pi  L^2 r_h \left(\frac{Q}{r_h^2}+1\right)^{\frac{3}{6 \delta ^2+2}} \left(3
   \delta ^2 \left(Q+r_h^2\right)+r_h^2\right)}{\left(3 \delta ^2+1\right) \kappa 
   L^2 \left(Q+r_h^2\right)+2 r_h^2 \left(6 \delta ^2 \left(Q+r_h^2\right)-Q+2
   r_h^2\right) \left(\frac{Q}{r_h^2}+1\right)^{\frac{3}{3 \delta ^2+1}}}.
    \end{aligned}
 \end{equation}
By studying how the ratio $r_h/r_0$ varies with $\tau/r_0$ for $\kappa=1$ and $\delta=\delta_c$, we can better understand their interdependence.
 We analyze the zeros of $\phi^{r_h}$ (see Fig. \ref{fig4}) and the unit vector field $n$ (see Fig. \ref{fig5}), specifically for $\tau = 200r_0^2$ and $\kappa=1$.
For the specified values of $L\sqrt{r_0^{1-d}}$ and $Q(\dfrac{r_h}{r_0})^{d-2}$, Fig.~\ref{fig4} reveals a generation point at $\tau/r_0^{d-2}=16.73$ and an annihilation point at $\tau/r_0^{d-2}=243.34$.
By calculating the winding number ($W$) for the three black hole branches shown in Fig.~\ref{fig5}, we observe a behavior that aligns with the new classifications introduced in \cite{wu2025novel}. Specifically, the $\textbf{AdS}_5$ black hole falls into the new thermodynamic topological class $W^{0-\leftrightarrow 1+}$, characterized by a winding number of $1$.
\subsection{ $\textbf{AdS}_6$ black holes}
\begin{figure}[H]
        \centering
        \includegraphics[width=0.5\linewidth]{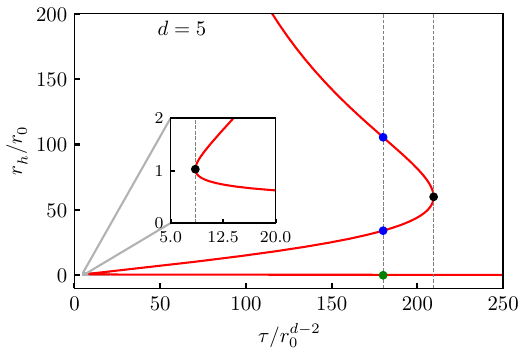}
        \caption{Zero points of the vector $\phi^{r_h}$ shown in the $r_h-\tau$ plane.}
        \label{fig6}
    \end{figure}

\begin{figure}[H]
    \centering
    \includegraphics[width=0.47\linewidth]{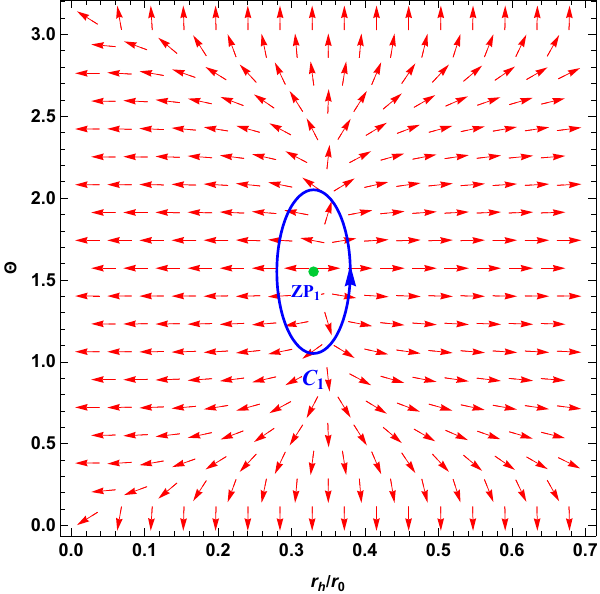}~~~
    \includegraphics[width=0.465\linewidth]{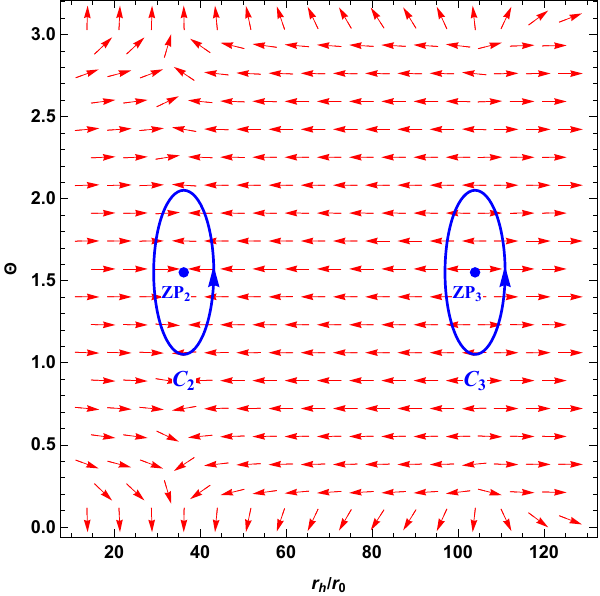}
    \caption{The unite vector field $n$ with $\tau=180r_0$ and $\kappa=1$.
    The zero points (ZPs) marked with green dot and blue dots at $(r_h,\Theta)=(0.33,\dfrac{\pi}{2})$ , $(r_h,\Theta)=(36.20,\dfrac{\pi}{2})$ and $(r_h,\Theta)=(104,\dfrac{\pi}{2})$, for $\text{ZP}_1$, $\text{ZP}_2$ and $\text{ZP}_3$, respectively.}
    \label{fig7}
\end{figure}

The key parameters for an $\textbf{AdS}_6$ black hole, such as its mass and entropy, which are necessary for computing the free energy $\mathcal{F}$, are provided below:
\begin{equation}
    \begin{aligned}
        \mathcal{E}&=\frac{4}{5} \kappa  \left(\frac{8 Q}{\left(6 \delta ^2+1\right) L^2}+3 r_h^3\right)+\frac{4
   r_h^5 \left(\frac{Q}{r_h^3}+1\right)^{\frac{8}{18 \delta ^2+3}}}{L^2},\\
        S&=4 \pi  r_h^4 \left(\frac{Q}{r_h^3}+1\right){}^{\frac{4}{18 \delta ^2+3}}.
    \end{aligned}
\end{equation}
Also, we have:
\begin{equation}
    \mathcal{F}=\frac{4}{5} \kappa  \left(\frac{8 Q}{\left(6 \delta ^2+1\right) L^2}+3 r_h^3\right)+\frac{4
   r_h^5 \left(\frac{Q}{r_h^3}+1\right)^{\frac{8}{18 \delta ^2+3}}}{L^2}-\frac{4 \pi  r_h^4 \left(\frac{Q}{r_h^3}+1\right){}^{\frac{4}{18 \delta ^2+3}}}{\tau }.
\end{equation}
We can easily determine the components of vector $\phi$:
\begin{equation}
     \begin{aligned}
     \phi^{r_h}&=\frac{4 r_h^4 \left(30 \delta ^2 \left(Q+r_h^3\right)-3 Q+5 r_h^3\right)
   \left(\frac{Q}{r_h^3}+1\right)^{\frac{8}{18 \delta ^2+3}}}{\left(6 \delta ^2+1\right)
   L^2 \left(Q+r_h^3\right)}-\frac{16 \pi  r_h^3 \left(6 \delta ^2
   \left(Q+r_h^3\right)+r_h^3\right)
   \left(\frac{Q}{r_h^3}+1\right)^{\frac{4}{18 \delta ^2+3}}}{\left(6 \delta ^2+1\right)
   \tau  \left(Q+r_h^3\right)}\\&+\frac{36 \kappa  r_h^2}{5},
   \\ \phi^{\Theta}&=-\cot{\Theta}\csc{\Theta},
   \end{aligned}
 \end{equation}
 Solving the equation $\phi^{r_h}=0$ yields a curve in the $r_h-\tau$ plane. For four-dimensional Gubser-Rocha models, this results in:
  \begin{equation}
    \begin{aligned}
        \tau=\frac{20 \pi  L^2 r_h \left(\frac{Q}{r_h^3}+1\right)^{\frac{4}{18 \delta ^2+3}}
   \left(6 \delta ^2 \left(Q+r_h^3\right)+r_h^3\right)}{9 \left(6 \delta ^2+1\right)
   \kappa  L^2 \left(Q+r_h^3\right)+5 r_h^2 \left(30 \delta ^2
   \left(Q+r_h^3\right)-3 Q+5 r_h^3\right)
   \left(\frac{Q}{r_h^3}+1\right)^{\frac{8}{18 \delta ^2+3}}}.
    \end{aligned}
 \end{equation}
To understand the relations between $r_h/r_0$ and $\tau/r_0$, we'll analyze their behavior across various $\kappa$ values ($0, 1, -1$). We'll also investigate the roots of $\phi^{r_h}$ (Fig.~\ref{fig6}) and the unit vector field $n$ (Fig.~\ref{fig7}) ,with specific values of $\tau=180r_0^3$ and $\kappa=1$.
It is important to observe that, for the given values of $L\sqrt{r_0^{1-d}}$ and $Q\left(\dfrac{r_h}{r_0}\right)^{d-2}$, a generation point and an annihilation point appear in Fig.~\ref{fig6} at $\tau/r_0^{d-2} = 8.53$ and $209.44$, respectively.
By evaluating the winding number $W$ for the three black hole branches shown in Fig.~\ref{fig6}, it becomes evident that this behavior aligns with the new classification scheme proposed in \cite{wu2025novel}.
The $\textbf{AdS}_6$ black hole falls into the newly defined thermodynamic topological category $W^{0-\leftrightarrow 1+}$, characterized by a winding number of 1.
\begin{figure}[H]
        \centering
        \includegraphics[width=0.6\linewidth]{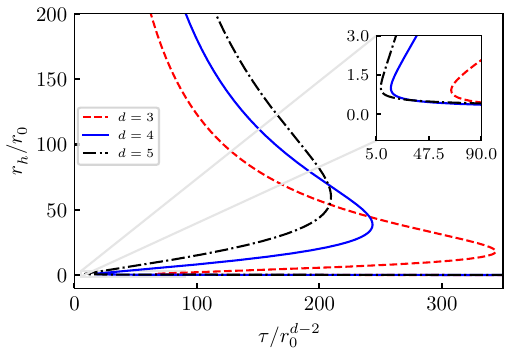}
        \caption{This figure illustrates the change in the graph's behavior at the critical value $\delta=\delta_c$ for dimensions 4, 5, and 6.}
        \label{fig8}
    \end{figure}
    At the critical value, $\delta = \delta_c$, the graph's behavior in the $r_h-\tau$ plane changes significantly for dimensions 4, 5, and 6 (Fig.~\ref{fig8}). The differences are particularly noticeable at $\tau_c$.
\section{Conclusions and outlook}\label{sec4}
In this work, we have explored the thermodynamic topology of a broad class of charged, asymptotically AdS black holes arising from Einstein-Maxwell-Dilaton (EMD) theories. By analytically investigating solutions in dimensions 4, 5, and 6, we have extended the framework of thermodynamic topology to new ground, highlighting the intricate connections between dimensionality, scalar couplings, and black hole thermodynamic behavior.

Our main findings emphasize the role of the dilaton coupling parameter $\delta$, particularly its critical value $\delta = \delta_c$, in governing the topological structure of black hole thermodynamics. For black holes in the Gubser-Rocha class and its generalizations, we observed transitions between topological classes as $\delta$ varied. Notably, in all analyzed dimensions, we found that black hole configurations exhibit zero points of the free energy gradient vector field that align with the newly proposed thermodynamic topological class $W^{0- \leftrightarrow 1+}$, characterized by a winding number $W = 1$. This classification reflects a novel sequence of stable and unstable black hole branches as the temperature changes, not previously captured by the original four classes introduced in the literature.

Our analysis confirms that the winding number serves as a powerful global topological invariant for classifying black hole phases. Furthermore, the generation and annihilation of topological defects (zero points of the vector field) with varying temperature reinforces the robustness of thermodynamic topology as a diagnostic tool in gravitational theories.

These findings suggest a more profound universality in black hole thermodynamics, rooted in topological and geometric considerations. They also underscore the capacity of string-theoretic and higher-dimensional models, such as the Gubser-Rocha framework, to illuminate fundamental aspects of black hole physics.

Looking forward, there are several promising directions for further research. These include extending the current analysis to rotating or dyonic black hole solutions, exploring time-dependent or dynamical spacetimes, and probing the implications of thermodynamic topology in holography and quantum gravity. Additionally, understanding the interplay between topological phase transitions and dual field theory phenomena within the AdS/CFT correspondence remains an intriguing and open area.

Ultimately, thermodynamic topology not only provides a novel language for classifying black holes but also bridges thermodynamics, geometry, and quantum field theory, offering fresh insights into the fundamental nature of gravity.

\section*{Acknowledgment}
The authors thank Pavan K. Yerra and Bin Chen for their interest in this work and the fruitful discussions. The work of H.B.-A. was conducted as part of the PostDoc Program on {\it Exploring TT-bar Deformations: Quantum Field Theory and Applications}, sponsored by Ningbo University. This research was partly supported by NSFC Grant No.12475053 and 12235016.
\bibliography{ref}
   \end{document}